\documentclass[aps,pre,showpacs,twocolumn,floatfix,superscriptaddress]{revtex4}

\usepackage{graphicx}
\usepackage{amsmath}

\usepackage{xspace} 
\newcommand{\nonmotile}{A$^-$S$^-$Frz$^-$\xspace}
\newcommand{\amotile}{A$^+$S$^-$Frz$^-$\xspace}
\newcommand{\asmotile}{A$^+$S$^+$Frz$^-$\xspace}
\newcommand{\wildtype}{A$^+$S$^+$Frz$^+$\xspace}

\usepackage{color}

\begin{document}

\title{Pattern formation mechanisms in motility mutants of \textit{Myxococcus xanthus} }

\date{\today}



\author{J\"orn Starru\ss \footnote{These authors have contributed equally to this work.}}
\affiliation{These authors have contributed equally to this work.} 
 \affiliation{Center for Information Services and High Performance Computing (ZIH), Technische Universit\"at Dresden, Zellescher Weg 12, D-01069 Dresden, Germany}
\author{Fernando Peruani \footnote{These authors have contributed equally to this work.} \footnote{Corresponding author: peruani$@$unice.fr}}
\affiliation{These authors have contributed equally to this work.} 
\affiliation{Laboratoire J.A. Dieudonn{\'e}, Universit{\'e} de Nice Sophia Antipolis, UMR 7351  CNRS , Parc Valrose, F-06108 Nice Cedex 02, France} 
\author{Vladimir Jakovljevic}
 \affiliation{Max Planck Institute for Terrestrial Microbiology, Karl-von-Frisch Stra{\ss}e 10, D-35043 Marburg, Germany}
 \author{Lotte S{\o}gaard-Andersen}
\affiliation{Max Planck Institute for Terrestrial Microbiology, Karl-von-Frisch Stra{\ss}e 10, D-35043 Marburg, Germany}
 \author{Andreas Deutsch}
\affiliation{Center for Information Services and High Performance Computing (ZIH), Technische Universit\"at Dresden, Zellescher Weg 12, D-01069 Dresden, Germany}
 \author{Markus B\"ar}
\affiliation{Physikalisch-Technische Bundesanstalt, Abbestra{\ss}e 2-12, 10587 Berlin, Germany}


\begin{abstract}

Formation of spatial patterns of cells is a recurring theme in biology and often depends on regulated cell motility. Motility of the rod-shaped cells of the bacterium {\it Myxococcus xanthus} depends on two motility machineries, type IV pili (giving rise S-motility) and the gliding motility apparatus (giving rise to A-motility).
Cell motility is regulated by occasional reversals.
Moving {\it M. xanthus} cells can organize into spreading colonies or spore-filled fruiting bodies depending on their nutritional status. 
To ultimately understand these two pattern formation processes and the contributions by the two motility machineries, as well as the cell reversal machinery, 
we analyze spatial self-organization in three {\it M. xanthus} strains: i) a mutant that moves unidirectionally without reversing by the A-motility system only, ii) a unidirectional mutant 
that is also equipped with the S-motility system, and iii) the wild-type that, in addition to the two motility systems, occasionally reverses its direction of movement. 
The mutant moving by means of the A-engine illustrates that collective motion in the form of large moving clusters can arise in gliding bacteria due to steric interactions of the rod-shaped cells, without the need of invoking any biochemical signal regulation. 
The two-engine strain mutant reveals that the same phenomenon emerges when both motility systems are present, and as long as cells exhibit unidirectional motion only. 
From the study of these two strains, we conclude that  unidirectional cell motion induces the formation of large moving clusters at low and intermediate densities, 
while it results into vortex formation at very high densities.
These findings are consistent with what is known from self-propelled rod models which strongly suggests that 
the  combined effect of self-propulsion and volume exclusion interactions is the  pattern formation mechanism leading to the observed phenomena. 
On the other hand, we learn that when cells occasionally reverse their moving direction, as observed in the wild-type, cells form small but strongly elongated clusters 
and self-organize into a mesh-like structure at high enough densities. 
These results have been obtained from a careful analysis of the cluster statistics of ensembles of cells, and analyzed on the light of a coagulation Smoluchowski equation with fragmentation.  


\end{abstract}

\maketitle

\begin{figure*}
\centering\resizebox{14cm}{!}{\includegraphics{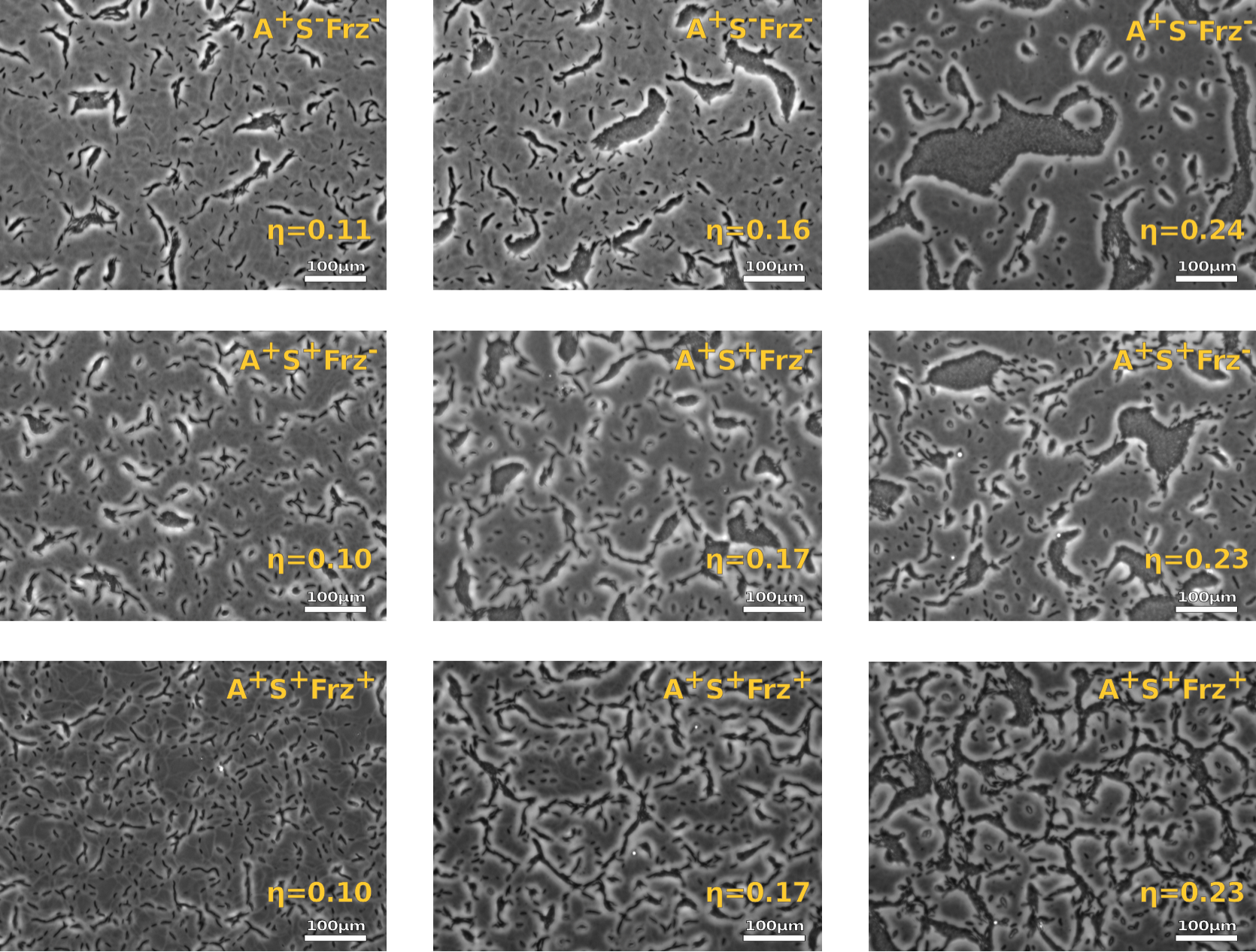}}
\caption{Pattern formation at various packing fractions $\eta$, 6 hs after spotting a drop of bacterial suspension on an agar surface. 
First row corresponds to the non-reversing \amotile mutant that moves only by means of the A-motility system (these three panels have been taken from~\cite{ourPRL}). 
Second row corresponds to the non-reversing \asmotile mutant that moves with both the A- and S-motility system.
%
%
At high cell densities, the mutants \amotile and \asmotile form large moving clusters that turn into vortices at sufficiently high packing fractions. 
Third row corresponds to the wild-type \wildtype strain that moves with both the A- and S-motility systems and cells are able to reverse their moving direction. The \wildtype mutant self-organizes into a mesh-like structure at high density.
}
\label{fig:severalStrains}
\end{figure*}

Formation of patterns of spatially organized cells is a recurring theme in biology. These processes often depend on regulation of cell motility. 
For instance, in metazoans it provides the basis for organ formation during embryogenesis and, in single-celled eukaryotes such as {\it Dictyostelium discoideum}, it is essential for the formation of fruiting bodies. 
In bacteria, regulated cell motility is essential for the colonization of diverse habitats  as well as for the formation of multicellular structures such as biofilms and fruiting bodies.
These pattern formation processes can be self-organized. For example, {\it Dictyostelium discoideum}~\cite{r10,r20} regulates cell aggregation and multicellular organization by secreting and sensing the diffusive signal cAMP. In {\it Myxococcus xanthus}, on the other hand, rippling patterns and the highly complex cellular reorganization 
leading to fruiting body formation are controlled by a non-diffusing signal, the C-signal~\cite{r30}. 
Interestingly, collective effects and self-organization can also occur, to a certain extent, in the absence of an explicit signaling mechanism. 
For instance, hydrodynamic interactions can induce large-scale coherent motion of swimming cells, as recently observed in {\it Bacillus subtilis}\cite{r35, r1},  
and a density-dependent diffusivity can lead to aggregation patterns as recently suggested to occur in {\it Escherichia coli} and {\it Salmonella typhimurium}~\cite{cates2012}. 

{\it M. xanthus} is a gliding bacterium that has been used as a model system to study pattern formation~\cite{r40}, bacterial social behaviour~\cite{r50}, and motility~\cite{r65}. 
The rod-shaped cells of the bacterium {\it M. xanthus} move on surfaces in the direction of their long axis using two motility machineries, type IV pili, 
which requires cell-to-cell contact for its activityÊ beacuse it is stimulated by exopolysaccharides on neighboruing cells~\cite{r70} (giving rise S-motility), and the 
gliding motility apparatus that allows cells to move in isolation~\cite{r80} (giving rise to A-motility). 
Force generation by the A-motility system has been suggested to rely on either slime secretion from the lagging pole~\cite{r112}, 
or on focal adhesion complexes distributed along the cell~\cite{r116}. 
Cells occasionally reverse their gliding direction with an average frequency of about once per 10 min and the reversal frequency is controlled by the {\it Frz} chemosensory system~\cite{r90}. 
In the presence of nutrients, {\it M. xanthus} cells form coordinately spreading colonies. 
Upon depletion of nutrients {\it M. xanthus} cells initiate a complex developmental program that culminates in the formation of spore-filled fruiting bodies. 
Both motility systems as well as reversals are required for the two cellular patterns to form, i.e., spreading colonies and fruiting bodies. 
It is currently not known how the reversal frequency is regulated except that cell-cell contacts may induce C-signal exchange which is supposed to stimulate reversals during rippling  
and to inhibit reversals during aggregation. 
During fruiting body formation the reversal frequency decreases up to a point where cell movements become nearly unidirectional~\cite{r100} and 
cells start to display collective motion with the formation of large clusters in which cells are aligned in parallel making side-to-side as well as 
head-to-tail contacts and move in the same direction~\cite{r110}. 
Eventually cells start to aggregate. Aggregation centers often resemble at their initial phase a cell vortex.

Here, we aim at understanding myxobacterial pattern formation processes, particularly the contributions by the two motility machineries as well as the cell reversal machinery to the spatial  organization of the cells. 
We study the role of  steric interactions, cell adhesion, and reversal frequency on the collective dynamics.  
The question for us is not ``why" cells exhibit a given collective behavior but ``how" they do it. 
In order to identify the role of the two motility machineries and cell reversal machinery, we follow a bottom-up strategy  by looking at the collective dynamics of different mutants of increasing complexity. 
We analyze three {\it M. xanthus} strains: i) a mutant that moves unidirectionally without reversing by the A-motility system only --  mutant that has been previously studied by us in~\cite{ourPRL} --, ii) a unidirectional mutant 
that is also equipped with the S-motility system, and iii) the wild-type that, in addition to the two motility systems, occasionally reverses its direction of movement. 
We characterize the macroscopic patterns mainly through the cluster statistics, in particular in terms of  cluster size and shape. 
We observe that the mutant moving by means of the A-engine only displays  collective motion in the form of large moving clusters. 
The study of its cluster size distribution reveals that  above a given density, clusters  can be arbitrary large~\cite{ourPRL}. 
Here, we show in addition that there is a non trivial scaling of cluster perimeter with cluster size which indicates that the clustering process is neither (fully) random nor as in (equilibrium) liquid-vapor drops~\cite{staffer}. 
We also find that at high densities the collective dynamics changes and cells organize into vortices.
%
%
The study of the two-engine strain mutant reveals the same phenomenology for  these bacteria: collective motion in the form of large moving clusters, a critical density above which cluster can be arbitrarily large,  a non-trivial scaling of cluster perimeter with cluster size, and vortex formation at high densities.
%
%
From the comparison of these two strains, we conclude that  unidirectional cell motion induces the formation of large moving clusters at low and intermediate densities,   
while it results into vortex formation at very high densities, see first and second row of panels in Fig.~\ref{fig:severalStrains}.
Interestingly, similar collective dynamics has been observed in self-propelled rod models~\cite{r32}, a fact that strongly suggests that 
the  combined effect of self-propulsion and volume exclusion interactions is the  pattern formation mechanism leading to the observed phenomena.
%

The study of  wild-type cells indicates that cell reversal weakens clustering. Wild-type cells exhibit exponential cluster size distributions at low and intermediate densities, while  
the scaling of the cluster perimeter with cluster size indicates that clusters are strongly elongated. 
At high densities, we find that reversing wild-type cells self-organize into a mesh-like structure, see bottom row in Fig.~\ref{fig:severalStrains}.

Wild-type cells, as commented above, exhibit  a large variety of self-organized patterns depending on the environmental condition. 
Our results suggest that by only switching on and off  the reversal,  cells can modify dramatically their collective behavior, with the suppression of cell reversal leading 
to collective motion in the form of moving clusters and vortex formation at high densities. 
This observation is consistent with the observed decrease in reversal frequency in the wild-type upon nutrient depletion, which is followed by the formation of large moving clusters and aggregation of cells. 
Our findings indicate that these two processes can result from simple steric  interactions of the (non-reversing) rod-shaped cells, without the need of invoking any biochemical signal regulation.

The paper is organized as follows. In Sec.ÊI.a, we focus on the spatial self-organization of purely A-motile cells in the absence of cell reversals. 
The effects induced by the S-motility engine, which include increased cell adhesion, are studied in Sec.ÊI.b, while those due to cell reversals in Sec.ÊI.c. 
In Sec.ÊII we discuss which collective effects are expected in self-propelled rod models, and interpret the cluster statistics results observed in the experiments 
in the light of a simple cluster formation theory. We summarize all the results in Sec. III, where we also discuss the implications of the reported findings. 

\begin{figure*}
\centering\resizebox{16cm}{!}{\rotatebox{0}{\includegraphics{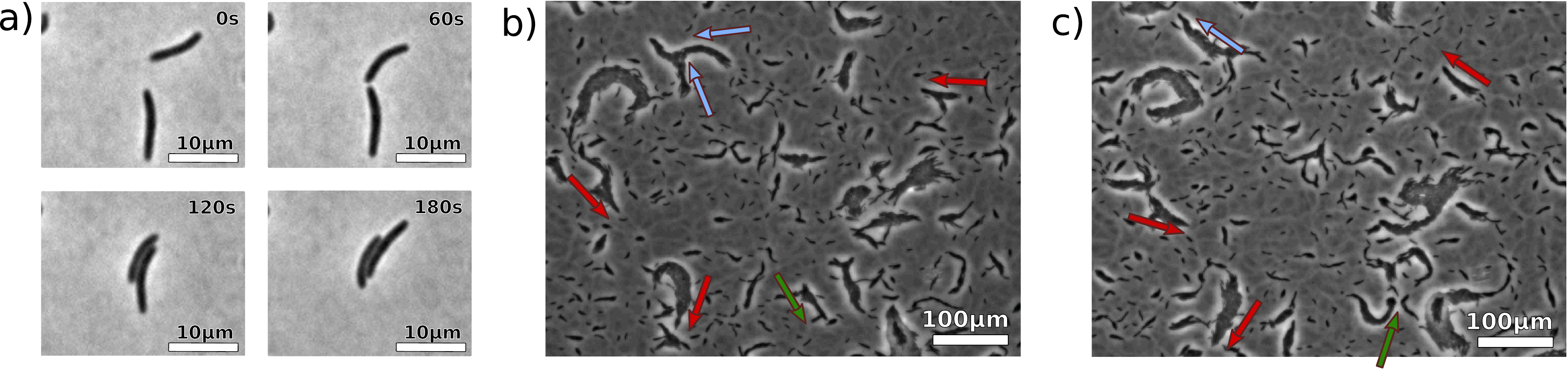}}}
\caption{(a) Collisions among {\it M. xanthus} lead to an effective (local) alignment. (b) and (c) show that a local alignment leads to formation of moving clusters; arrows indicate the cluster moving direction. 
Time interval between (b) and (c) is 15 min, snapshots correspond to \amotile cells at packing fraction $\eta =0.11$. Panels taken from~\cite{ourPRL}.}
\label{fig:localAlignment}
\end{figure*}

\section{Cluster statistics}

\subsection{ A-motile non-reversing cells}

We start out with the simple mutant \amotile that only moves by means of the A-motility system and which is unable to reverse due to an insertion in the {\it frz} gene cluster (see {\it Material and Methods} for more details about how the strain was generated). 
This mutant is unable to assemble type IV pili due to deletion of the pilA gene, which encodes the type IV pili subunit,  and therefore the S-motility system is non-functional in this mutant. 
This mutant exhibits relatively weak cell-cell adhesion due to the lack of type IV pili and the reduced accumulation of exopolysaccharides. 
This mutant is labeled \amotile to indicate that A-motility engine is on,  the S-motility engine is off and the Frz system, i.e., cell reversal is dramatically reduced.   
Control experiments showed that these mutants  have a reversal period $\gg100$ min whereas the isogenic Frz$^+$ strain reversed with a mean reversal  period of $\sim 10$ min. 
In~\cite{ourPRL}, we showed that this mutant exhibits a transition to a collective motion phase at high enough densities by analyzing the dependency of cluster size distribution with the packing fraction. Here,  we characterize in addition the cluster shape, and show that at  densities higher than the one studied in~\cite{ourPRL}, giant clusters turn into vortices.  

Experiments were performed by spotting a drop of cell suspension of the desired density on an agar surface to subsequently monitor the evolution of cell arrangements by taking snapshots of the bacterial colony 
every 30 min for a total of 8 hrs. 
Experiments with cells gliding in isolation indicate an average velocity of $v=3.10 \pm 0.35$ $\mu$m/min, an average width of about $W=0.7$ $\mu$m and an 
average length of $L=6.3$ $\mu m$. This results in a mean aspect ratio of $\kappa=L/W=8.9 \pm 1.95$ and a cell covering an average area $a=4.4$ $\mu m^2$. 

We found that under these conditions  cells organized over time into moving clusters. 
Time-lapse recordings showed that collisions of cells lead to effective  alignment (Fig.~\ref{fig:localAlignment}a). 
When the interaction is such that cells end up parallel to each other and move in the same direction, they migrate together for a long time 
(typically $>15$ min). 
Eventually, successive collisions allow a small initial cluster to grow in size, Fig.~\ref{fig:localAlignment}. 
In the individual clusters, cells are aligned in parallel to each other and arranged in a head-to-tail manner, as previously described~\cite{r130}. 
In a cluster, cells move in the same direction. 
Cluster-cluster collision typically leads to cluster fusion, whereas splitting and break-up of clusters
 rarely occur. 
On the other hand,  individual cells on the border of a cluster often spontaneously escape from the cluster. 
These two effects, cluster growth due to cluster-cluster collision and cluster shrinkage, mainly due to cells escaping from the cluster 
boundary, compete and give rise to a characteristic cluster size distribution (CSD). 

\begin{figure*}
\centering\resizebox{12 cm}{!}{\rotatebox{0}{\includegraphics{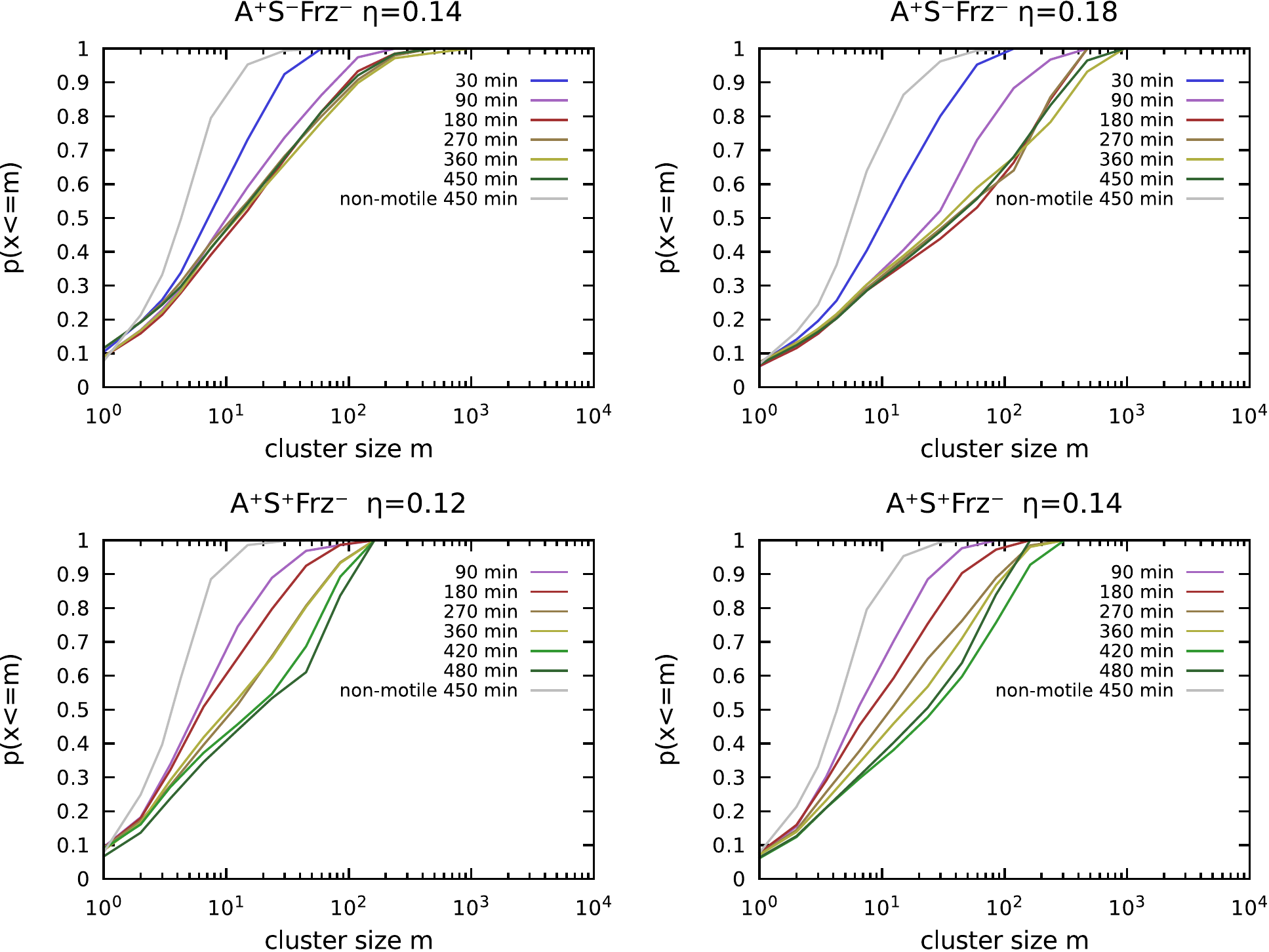}}}
\caption{Time convergence towards a steady state. 
The figure compares the cumulative cluster size distribution (CCSD),  defined as $p(x<=m)$, at various time points for \amotile and  \asmotile cells at two different packing fractions. 
The CCSD is less noisy than the CSD and the comparison at various time points becomes possible. 
The first row, corresponding to \amotile cells, indicates that the cluster statistics quickly converges to a steady state. 
The time convergence for  \asmotile cells, second row, also occurs, though the phenomenon is less evident. 
Each panel shows, as reference, the CCSD obtained with control experiments of non-motile cells. 
The comparison indicates that cell motility promotes undoubtedly the formation of large clusters.  
}
\label{fig:convergenceTime}
\end{figure*}

The CSD - $p(m,t)$ - indicates the probability of a bacterium to be in a cluster of size $m$ at time $t$. 
Note that along the text, the term CSD always refers to this definition. 
Often times the cluster size distribution is alternatively determined as the number $n_m(t)$ of clusters of size $m$ at time $t$. 
There is a simple relation between these two definitions: $p_m(t) \propto m\,n_m(t)$. 
In experiments we have observed that  the CSD mainly depends on the packing fraction $\eta$, where  $\eta = \rho\,a$, with $\rho$ the
 (two-dimensional) cell density and $a$ the average covering area of a bacterium given above.
Hence, for all snapshots first the packing fraction was determined. 
Then, images with similar packing fraction $\eta$ were compared and the CSD was reconstructed  by determining the CSD for all images within a 
finite interval of the packing fraction. 
Very importantly, we find that the CSD $p(m,t)$ reaches a steady state $p(m)$  after some transient time, as shown in Fig. \ref{fig:convergenceTime}. 
We conclude that  the clustering process evolves towards a dynamic equilibrium, where the process of  formation of cell clusters of a given size 
is balanced by events in which clusters of this size  disappear by either fusing with other clusters or by loosing individual cells from their boundary. 

\begin{figure*}
\centering\resizebox{18cm}{!}{\rotatebox{0}{\includegraphics{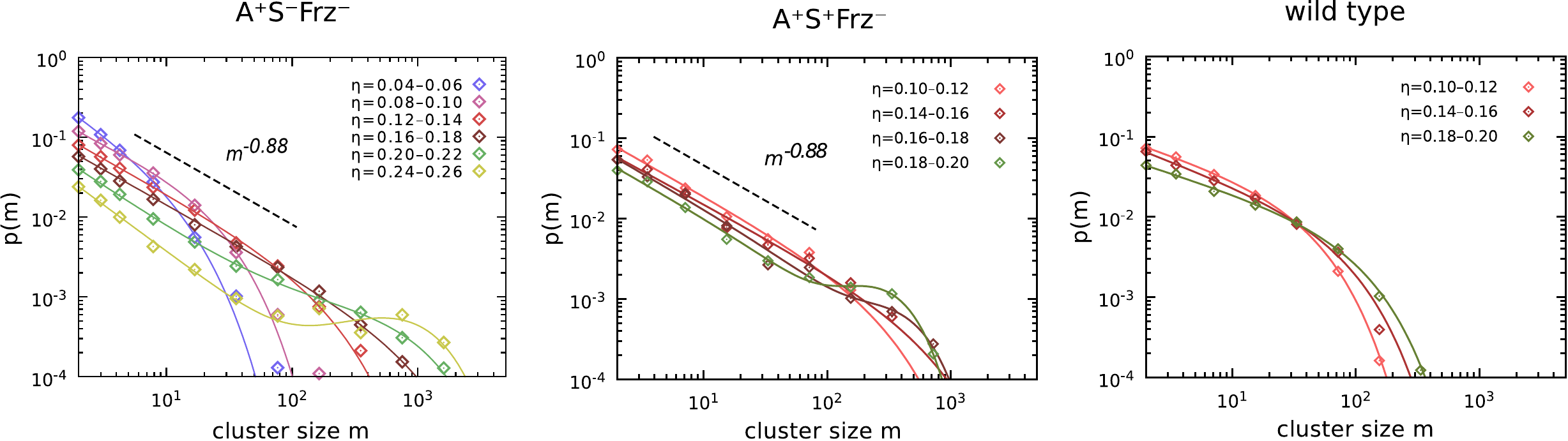}}}
\caption{Asymptotic cluster size distribution ($t=450$ min) at various packing fractions $\eta$ for non-reversing mutants \amotile and \asmotile and \wildtype. 
The three strains exhibit a cluster dynamics that evolves towards a steady cluster size distribution which is function of the cell packing fraction. 
The cluster size distribution (CSD) for \amotile and \asmotile  cells exhibits a qualitative change at a critical packing fraction $\eta_c \sim 0.17$; for $\eta > \eta_c$ the CSD is no longer monotonically decreasing distribution and a peak at large cluster sizes emerges. At the critical point,  $p(m) \propto m^{-\xi} $, with $\xi \sim 0.88$. 
%
The CSD of the densities examined can always be approximated by a power-law with an exponential cut-off. 
Reversing, fully motile  \wildtype cells (wild-type) display an asymptotic CSD which is for all packing fractions $\eta < \eta^*$ exponential. For  $\eta > \eta^*$, clusters connect such that cells form a mesh-like structure as shown in Fig. \ref{fig:severalStrains}.  
}
\label{fig:CSD_mutants}
\end{figure*}

The steady-state CSD $p(m)$ strongly depends on the packing fraction $\eta$, with more and more cells moving in larger clusters for increasing packing 
fraction $\eta$.
This is evident in Fig.~\ref{fig:CSD_mutants}, where we observe that at small values of $\eta$, $p(m)$ exhibits a monotonic sharp decay with $m$, 
while at large $\eta$ values, $p(m)$ is non-monotonic, with an additional peak at large cluster sizes. 
The solid curves in Fig.~\ref{fig:CSD_mutants} are fitted to the raw data by using phenomenological functional forms  described in the next section.  
The CSD here was determined at a fixed time (450 minutes) after the beginning of each experiment;  control experiments at other times 
(360 minutes, 480 minutes) revealed practically identical behavior. 
We interpret the presence of a peak at large values of $m$ at bigger values of the  packing fractions as the emergence of collective motion 
resulting in formation of large clusters of bacteria moving in a coordinated fashion. 
The clustering transition is evident by the functional change displayed by $p(m)$, monotonically decreasing with $m$ for small values of $\eta$, 
while exhibiting a local maximum at large $\eta$  values. 
At a critical value $\eta_c = 0.17 \pm 0.02$ that separates different regimes of behavior, the CSD can be approximated by
 $p(m) \propto m^{-\gamma_0}$, with $\gamma_0 = 0.88 \pm 0.07$. 
%
Control experiments with non-motile cells do not exhibit a power-law behavior in the CSD. 
For more details, we refer to reader to~\cite{ourPRL}.
Hence, we conclude that without active motion of cells no comparable transition to clustering occurs. 
In other words, active motion is required for the dynamical self-assembly of cells.

Now, we turn our attention to the cluster shape, in particular to the scaling of the cluster perimeter $\Pi(m)$ with the cluster size $m$.  
This kind of information can help us to realize how adhesive cells are and which role adhesion plays in the clustering process. 
If there is surface tension, then cluster should tend to minimize their surface, and they should be round, as observed in liquid-vapor drops~\cite{staffer}. 
On the other hand, if surface tension is negligible, cluster can be very elongated object, with most of the cells on the cluster boundary, and the cluster perimeter is proportional to cluster size.  
We assume that $\Pi(m) \propto m^{\omega}$, where $m$ denotes the area of the cluster. Thus it is clear that perimeter exponent $\omega$ should be $0.5$ for round clusters. 
This would be the case for very adhesive cells exhibiting random movements. 
If clusters are extremely elongated, then $\omega=1$. 
We notice that $\omega=1$ would correspond also  a fully random process as observed in percolation theory~\cite{staffer}. 
In short, the exponent $\omega$ is then such that $0.5 \leq \omega \leq 1$. 
Fig. \ref{fig:clusterShape} shows that  for \amotile cells $\omega=0.60 \pm 0.03$, which indicates that the clustering process is non trivial that it neither fully random nor dominated by surface tension, 
see also Fig. \ref{fig:severalStrains}. 
The scaling of $\Pi(m)$ with $m$ plays a central role in the clustering theory we discuss below, where the relation between cluster size statistics and cluster perimeter statistics will be discussed in detail.

As the density increases, typical above $\eta > 0.26$, cells do not  organize into large moving clusters,  
and giant clusters evolve into  vortices. 
These vortices are formed by one or several layers of rotating disks whose radii 
diminish the higher the disk is located in the z-direction. 
Fig. \ref{fig:vortex} shows a typical example of vortex formation; see the supplementary material for a movie and~\cite{vortex_movie} for a brief description of the movie.   
Notice that these vortices are not disordered aggregates of cells as suggested 
in~\cite{holmes2010}. 
Given the fact that vortices are multilayered  structures, phase contrast imaging can only provide limited information regarding the actual cell arrangements inside vortices. 
A detailed study of vortices requires more sophisticated experimental techniques.

Interestingly, vortex formation has been also observed in other experimental ``self-propelled rod" systems as actin-myosin motility assays \cite{r400, sumino2012} 
as well as  in 2D suspensions of sperms~\cite{riedel2007}. In the later example, hydrodynamical interactions are supposed to induce the observed pattern, while in the former ones 
the role of hydrodynamic interactions is not well understood;  yet in both type of systems the vortex patterns correspond to vortex arrays. 
In myxobacteria, on the other hand, hydrodynamical effects can be neglected and vortices do not emerge in a lattice-like arrangement, but rather in a disorganized fashion. 
At a theoretical level, vortices has been found in active gel theory~\cite{Kruse2004, Elgeti2011}. 
Wether active gel vortices and those observed in {\it M. xanthus} mutants have the same microscopic origin is unclear, but 
certainly a possibility worth exploring.  

In summary, the finding of vortex formation in experiments with \amotile indicates that the S-motility system,  cell-to-cell signaling, and cell reversals are not required for the organization of cells into vortices. 

\subsection{A- and S-motile, non-reversing cells}

\begin{figure*}
\centering\resizebox{16 cm}{!}{\rotatebox{0}{\includegraphics{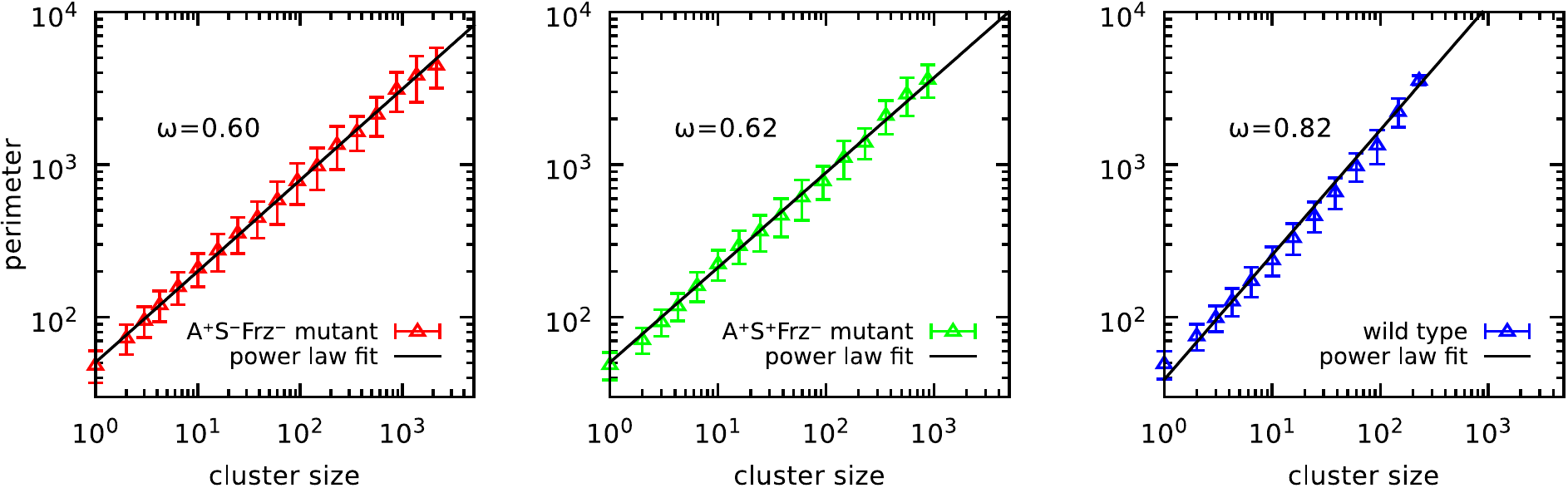}}}
\caption{Cluster perimeter $\pi(m)$ as function of the cluster size $m$ for three bacterial strains. 
The \amotile and \asmotile mutants exhibit roughly the same scaling $\pi(m) \propto m^{\omega}$, 
with $\omega \sim 0.62$ for the \amotile,  $\omega \sim 0.60$ for the \asmotile, suggesting that an increase in adhesion does not have a strong impact on the cluster shape. 
On the other hand, cell reversals lead to much more elongated clusters, as the scaling of the \wildtype cells indicates, with  $\omega \sim 0.82$. 
}
\label{fig:clusterShape}
\end{figure*}

We turn our attention to the next simplest mutant: \asmotile. 
These cells contain both motility engines found in the wild-type, while cell-reversals  are absent.  
The S-motility system depends on type IV pili~\cite{r70}. It allows cells to move in a contact-dependent manner, i.e. cells have to be in close proximity for S-motility to become active.
As previously reported~\cite{r70}, we find that \asmotile cells are more adhesive.
Our aim is to understand whether the S-motility engine affects the spatial self-organization of cells.
We performed the same analysis on \asmotile cells as described for \amotile cells and investigate cell densities close to the obtained critical density. 
Fig. \ref{fig:severalStrains} shows that at least at first glance the cluster statistics resembles that obtained with \amotile cells. 
This suggests that  the additional motility including its adhesion effect has no significant impact on the organization of cells within a cluster.
By looking in more detail on the clustering data  some subtle differences can be revealed. 
We observe that for all fixed packing fractions $\eta$, the CSD seems to evolve towards a steady state, Fig.~\ref{fig:convergenceTime}. 
However, the temporal convergence is slower than the one observed for \amotile cells.  
Assuming that CSD after 450 min from the beginning of the experiment is representative of the steady state CSD,  we show in Fig. \ref{fig:CSD_mutants} the asymptotic behavior of the CSD with packing fraction $\eta$. 
The CSDs of the packing fractions $\eta<0.18$  can be roughly approximated by a power-law, $p(m) \propto m^{-\gamma_0}$, with a critical exponent $\gamma_0$ consistent with the one obtained for  \amotile cells, i.e., $0.81 \leq \gamma_0 \leq 0.95$, see Fig. \ref{fig:CSD_mutants}. 
On the other hand, the data indicates that a local maximum, as the one described above for \amotile, emerge for $\eta \geq 0.18$, Fig. \ref{fig:CSD_mutants}. 
On the other hand, the cluster shape statistics shows that the scaling of the perimeter $\Pi$ with the cluster mass $m$ is again consistent with the one obtained for \asmotile cells with $\omega = 0.62 \pm 0.03$, see Fig. \ref{fig:clusterShape}. 
Finally, at sufficiently high densities, these cells also self-organize into vortices. 
%


%

%

\subsection{Wild-type and the effect of cell-reversal}

We applied the same analysis to the reversing \wildtype cells that move by means of both motility systems.
%

Fig. \ref{fig:severalStrains} shows that the spatial organization of wild-type cells is dramatically different from the one observed in the two mutants. 
Undoubtedly, cell-reversal has a strong impact on the macroscopic behavior of the colony. 
The CSD distribution after 450 min is exponential for all $\eta < 0.20$ as shown in Fig. \ref{fig:CSD_mutants}.
The net distance of cell movement is reduced due to cell reversals and cells can only form small clusters.
On the other side, clusters exhibit a  more elongated shape than those found in experiments with \amotile and \asmotile cells, as confirmed by  the scaling of the perimeter $\Pi(m)$ which is characterized by a very different exponent  $\omega = 0.82 \pm 0.03$, 
see Fig. \ref{fig:clusterShape}. 
The initial monodisperse phase, characterized by an exponential CSD and very elongated clusters, undergoes a transition at packing fractions larger than $0.26$. The new arrangement of cells percolates and the cells organize into a mesh-like structure, as shown in Fig. \ref{fig:severalStrains}.
\begin{figure}
\centering\resizebox{7 cm}{!}{\rotatebox{0}{\includegraphics{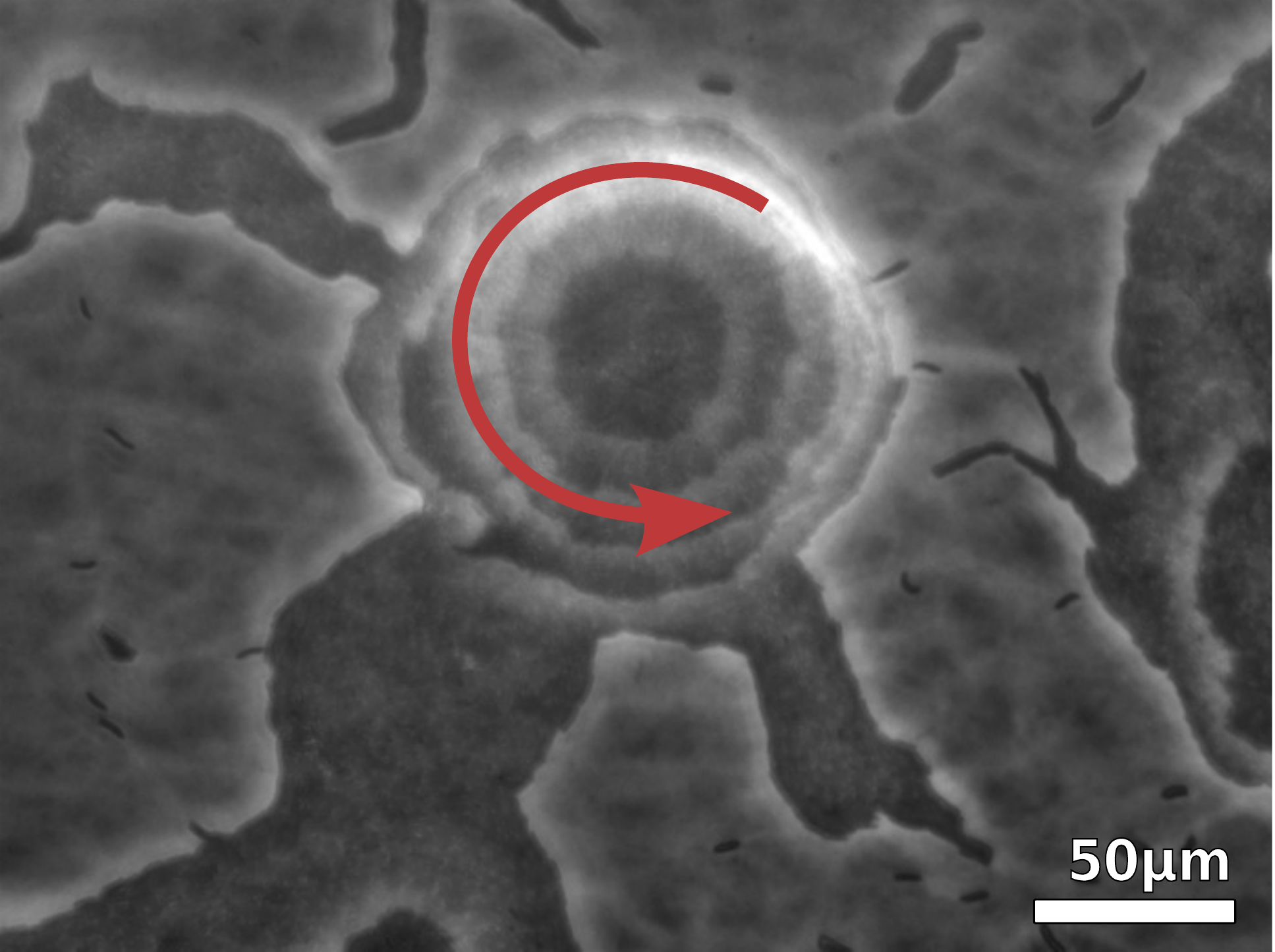}}}
\caption{Vortex formation (circular pattern in the center) in the non-reversing
mutant \amotile at high cell density ($\eta > 0.26$). 
The pattern consists of rotating stacked discs
of cells. These structure are observed in both, \amotile and \asmotile strains.}
\label{fig:vortex}
\end{figure}

\section{Kinetic model for the cluster statistics} \label{sec:smolu}


%
In the following, we outline  a generalized kinetic model for the
cluster-size distribution and compare it to the above experimental results. 
In particular, we want to relate the cluster size distribution data and the cluster shape statistics. 
The model equations are built on the well-established coagulation theory for colloidal particles originally suggested by Smoluchowski \cite{r164}, 
for an early review see also \cite{r166}. 
A similar phase transition (albeit with different exponents for the cluster-size distribution at criticality) was recently obtained in in a model 
for reversible polymerization representing a different generalization of the Smoluchowski  model \cite{r168}.

The model studied consists of a system of kinetic equations for the dynamics of the number $n_j(t)$ of clusters of size $j$ at time $t$. 
It was first proposed in~\cite{r32} to describe clustering in simulations of self-propelled rods. 
The individual cluster dynamics~\cite{peuani2010}, as well as the cluster-cluster dynamics~\cite{r32, peruani2010} are strongly simplified 
in this kinetic theory where 
the time evolution of the number $n_j(t)$ of clusters of size $j$  is simply given by:
\begin{eqnarray}
\dot{n}_{1}&=&2B_{2}n_{2}+\sum_{k=3}^{N}B_{k}n_{k}-\sum_{k=1}^{N-1}A_{k,1}n_{k}n_{1}
\nonumber \\
\dot{n}_{j}&=&B_{j+1}n_{j+1}-B_{j}n_{j}-\sum_{k=1}^{N-j}A_{k,j}n_{k}n_{j}
\nonumber \\
&&+\frac{1}{2}\sum_{k=1}^{j-1}A_{k,j-k}n_{k}n_{j-k} \ \quad \mbox{for} \quad j = 2, .....,N-1
\nonumber \\
\dot{n}_{N}&=&-B_{N}n_{N}+\frac{1}{2}\sum_{k=1}^{N-1}A_{k,N-k}n_{k}n_{N-k} \, , \label{eq:smolu}
\end{eqnarray}
where the dot denotes a time derivative and $N$ is  the total number of cells in the system. 
The cluster-size distribution  is then simply obtained from 
\begin{equation} \label{eq:pm}
p(m,t) = \frac{m\,n_m(t)}{N} \, .
\end{equation}
We have assumed that aggregation of cells occurs only due to  cluster-cluster collisions. 
Following earlier work \cite{r32, peruani2010}, the collision rate between clusters of mass $j$ and $k$ is defined by:  
\begin{eqnarray}\label{eq:rate_collision}
A_{j,k}= \frac{v_0  \sigma_0}{\delta}\left(\sqrt{j}+\sqrt{k}\right) \, ,
\end{eqnarray}
where $v_0$ represents the average speed of individual cells, $\sigma_0$ is
the average scattering cross section of a single cell which is assumed to be 
$\sigma_0 \approx L+W = \sqrt{a}(\sqrt{\kappa} + 1/\sqrt{\kappa})$ and $\delta$ is the total area  of the system. 
Eq. (\ref{eq:rate_collision})  assumes that cluster have a well-defined direction of motion, which means that the equation is not adequate to describe 
cluster-cluster coagulation in experiments with wild-type cells. 
This process competes with cluster fragmentation stemming from the escape of individual single cells from the cluster boundary. 
The fragmentation rate is given by the expression
\begin{eqnarray}\label{eq:rate_splitting}
B_{j} =  \frac{v_0 j^{\omega}}{R_0 L}= \frac{v_0 j^{\omega}}{R_0 \sqrt{a \kappa}} \, ,
\end{eqnarray}
where $R_0$ is a proportionality constant that is the only free parameter in the theory that is used to fix
 the critical value $\eta_c \propto R_0^{-1}$ at the same values as in the experiment. 
The exponent $\omega$ in the fragmentation rate has an important role: it represents the scaling between the cluster mass $m$ and the cluster
 perimeter $\Pi$, i.e., $\Pi \propto m^{\omega}$. 
If one assumes large clusters of approximately circular shape, then $\omega=1/2$; this special case has been previously studied in \cite{r32}.
If instead one considers that cells form elongated narrow clusters, where practically all cells are near the boundary,  then a choice of 
$\omega=1$ is appropriate. 
In practice, the value of $\omega$ will depend on the number $j$ of particles  in  the respective cluster. 
For simplicity, we study only the limiting cases $\omega = 1/2$ and $\omega = 1$ and compare the resulting cluster-size distribution to
 the experimental findings.  
%
%
%
According to the model, the exponent $\gamma$ only depends on the scaling of $\Pi(m)$, i.e., the exponent $\omega$, 
while the critical packing fraction $\eta_c$ is a non-universal quantity. 
%
%
The analysis of Eqs.~(\ref{eq:smolu}), performed by direct numerical integration using a fourth-order Runge-Kutta method,  reveals that  for $\eta\leq\eta_c$, the scaling 
of $p(m)$ takes the form:
\begin{equation}\label{eq:scaling}
p(m) \propto  m^{-\gamma_0} \exp(-m/m_0) \, , 
\end{equation}
while above it, i.e., for $\eta > \eta_c$, the scaling is:
\begin{equation}\label{eq:scaling_b}
p(m) \propto  m^{-\gamma_1} \exp(-m/m_1) + C m^{\gamma_2} \exp(-m/m_2) \, ,  
\end{equation}
with $\gamma_1,\,\gamma_2,\,m_1,\,m_2$ and $C$ constants that depend on $\eta$ and system size.  
Eq. (\ref{eq:scaling}) is the result of a system size study of  Eqs.~(\ref{eq:smolu}) at the critical point (not shown), while Eq. (\ref{eq:scaling_b}) is just an educated guess. 
Eqs. (\ref{eq:scaling}) and  (\ref{eq:scaling_b}) 
have  been used to fit the experimental data for the cluster size distributions in the different strains of myxobacteria shown in Fig.~\ref{fig:CSD_mutants}. 
 For $\eta<\eta_c$, for either \amotile and \asmotile cells we find using Eq.~(\ref{eq:scaling})  $\gamma_0 \in [0.80,0.95]$ and $m_0 \in [20, 1300]$ ($m_0 \sim 20$ for $\eta =0.04$ and $m_0 \sim 1300$ for $\eta =0.16$). 
Nevertheless, the critical exponent $\gamma_0$ has been estimated by the method explained in the Material and methods section, where $\gamma_0$ is found to be $\gamma_0 = 0.88 \pm 0.07$.  
For wild-type cells, the distribution is strongly dominated by an exponential tail. Using Eq.~(\ref{eq:scaling}) we find $\gamma_0 \in [0,0.63]$ and  $m_0 \in [10, 120]$.

\begin{figure*}
\centering\resizebox{14cm}{!}{\rotatebox{0}{\includegraphics{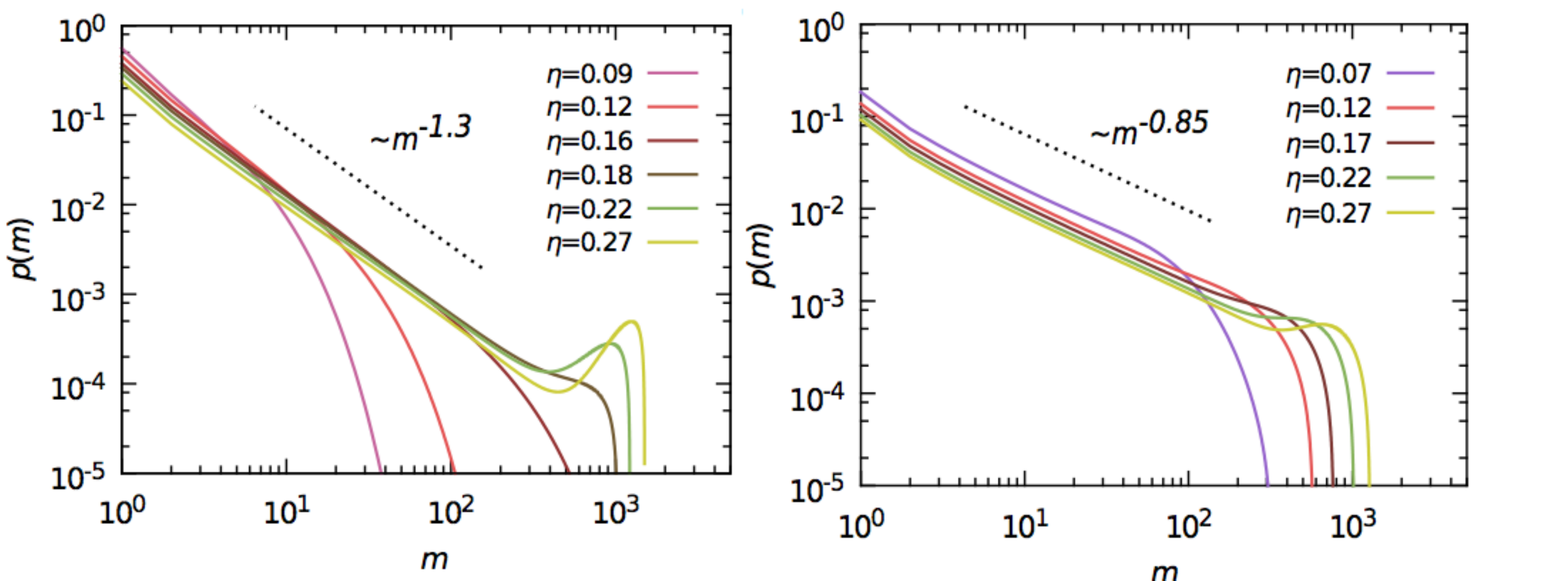}}}
\caption{Theoretical predictions for the cluster size distribution (CSD). 
The cluster size distribution from the kinetic model depends on the scaling of the fragmentation $B_m  \propto m^{-\omega}$ (see text). 
The figure shows the results for the  two limiting cases, in the left panel $\omega=0.5$ (and $R_0=0.58$)  corresponding to round clusters, while in right panel $\omega=1$ (and $R_0=11.9$), implying elongated clusters. 
Notice that in theory, as well as in experiments, the CSD can be well approximated by a power-law at the critical packing fraction $\eta_c$, while 
for $\eta > \eta_c$  a peak at large cluster sizes emerges, a signature of dynamic self-assembly into larger moving clusters. Other model parameters: $\kappa=9$, $a=4.4 \mu m^2$, and $\delta=699\times522 \mu m^2$}\label{fig:CSD_theory}
\end{figure*}

Through  Eq. (\ref{eq:smolu}), 
it can be shown that $m_0$ is a function of $\eta$ that increases as $\eta_c$ is approached from below as observed in Fig.~\ref{fig:CSD_mutants}. 
According to the kinetic model, the critical packing fraction $\eta_c$ is defined by  $p(m) \propto m^{-\gamma_0}$ at $\eta=\eta_c$ as long as 
$m$ is below the total number of cells $N$ in the system.
In contrast, for $\eta < \eta_c$ the function $p(m)$ clearly exhibits an exponentially decaying  tail at larger $m$, as observed in the experiments with \amotile and \asmotile cells, Fig.~\ref{fig:CSD_theory}. 
The theoretical CSD  $p(m,t)$ was obtained by numerical integration from an initial condition with $n_1 = N$ and $n_i = 0$ for $i \geq 2$. 
The values of the variables $n_i$ of Eq. (\ref{eq:smolu}) reached constant steady values after sufficiently large integration times.   
The steady state  $p(m)$  was found to depend only on the packing fraction $\eta$ for a given perimeter scaling characterized by $\omega$.  
For both values of $\omega$ studied, we find a transition from an exponentially decaying CSD, described by Eq. (\ref{eq:scaling}) for low densities, to a  non-monotonic CSD, described by Eq. (\ref{eq:scaling_b}), consisting of a power-law behavior for small cluster sizes and a  peak, local maximum, at large cluster sizes, see Fig.~\ref{fig:CSD_theory}. 
Upon closer inspection of the model results, one recovers distinctly different exponents for the different model assumptions regarding 
$\omega$: $\gamma_0 = 1.3$ for $\omega = 1/2$ and $\gamma_0 = 0.85$ for $\omega = 1$. 
%
%
%
%
%
Both choices of $\omega$ give reasonable qualitative agreement with the experimental data shown in Fig. \ref{fig:CSD_mutants} above, see Fig \ref{fig:CSD_theory}. 
Moreover, we find that the exponent of the cluster-size distribution is non-universal and depends sensitively on the choice of the fragmentation rate in Eq. (\ref{eq:rate_splitting}). 
We expect that changes in the collision rate for the cluster will have a similarly strong effect, as discussed below.  

The clustering model given by Eq. (\ref{eq:smolu})  allows to study the relationship between the perimeter scaling (characterized by an exponent $\omega$) and the cluster size distribution exponent $\gamma_0$. 
Eq. (\ref{eq:smolu})  also predicts  the existence of two CSDs, depending on the  the packing fraction $\eta$, i.e., Eqs. (\ref{eq:scaling}) and (\ref{eq:scaling_b}). 
%
%
These two predicted distributions  are found  in  experiments with  \amotile and \asmotile cells. 
For the wild type cells, only the CSD given by Eq. (\ref{eq:scaling}) is found. 
In this context, it is interesting to note that the results shown in Fig. \ref{fig:CSD_theory} imply that for $\omega =1$ one needs to assume a much 
lower fragmentation rate - indicated by a much larger value of the parameter $R_0$  than for $\omega = 0.5$ to obtain the same critical $\eta_c$.  
%
%
%
Beyond the apparent agreement between the CSD exhibited by Eq. (\ref{eq:smolu}) and the experimentally obtained CSDs for \amotile and \asmotile cells, there are important  differences.  
To obtain a critical  exponent $\gamma_0$ close to $0.88$,  $\omega$ has to be large, specifically, close to $1$,  
while the experimental measurements on $\Pi(m)$ revealed $\omega \sim 0.6$. 
%
%
There are several possibilities that could explain this discrepancy. 
For instance, 
%
the assumption that the cluster-cluster coagulation is proportional to square root of the cluster mass has to be revised.  
An estimation of the scaling of the effective scattering cross section of a cluster with its mass, as well as an accurate measurement of the functional dependency of  cluster speed with cluster mass would allow us to determine the correctness of Eq. (\ref{eq:smolu}). 
Unfortunately, such measurements are extremely difficult to obtained.  
Nevertheless, the apparent discrepancy suggests that a possible generalization of the presented clustering theory would include a modification of  Eq. (\ref{eq:rate_collision}).
%

\section{Discussion} \label{sec:conclu}

In order to identify the role of the two motility machineries as well as cell reversal machinery on the spacial collective dynamics of  {\it M. xanthus}, 
we have analyzed three bacterial strains: 
i) a mutant that moves unidirectionally without reversing by the A-motility system only, ii) a unidirectional mutant 
that is also equipped with the S-motility system, and iii) the wild-type that is equipped with the two motility systems and occasionally reverses its direction of movement. 
The study of the two non-reversing mutants revealed the same phenomenology. 
At low and intermediate densities, non-reversing cells displays  collective motion in the form of large moving clusters, with  a critical density above which clusters can be arbitrarily large. 
At the critical density, the two non-reversing strains exhibit a cluster size distribution characterized by roughly the same critical exponent $\gamma_0 \sim 0.88$. 
Even though the two-engine strain is supposed to be more adhesive than the single A-engine strain, we found  a similar non-trivial scaling of cluster perimeter with cluster size characterized by an exponent $\omega \sim 0.6$. 
This finding indirectly shows that the clustering process is, for both stains, neither fully random nor an equilibrium one as in liquid-vapor drops~\cite{staffer}. 
%
In order to connect the statistics on cluster size and cluster shape, we derived a Smoluchowski-coagulation theory with fragmentation, where we related the scaling of cluster perimeter with cluster size with the fragmentation kernel. 
The proposed theory allows us to understand the cluster formation process in absence of adhesion as a  dynamic self-assembly process. It predicts the existence of a steady state cluster size distribution which is function of the cell density and perimeter exponent $\omega$, and a functional change of the cluster size distribution above a critical density. 
In addition, the proposed theory predicts that the critical exponent $\gamma_0$  depends on the perimeter exponent $\omega$ only. 
In summary, the theoretical clustering model provides a qualitative descriptions consistent with the experimental measurements, and explains 
 why if the value of $\omega$ is similar for both strains, the value of  $\gamma_0$ has to be also similar. 
We observe that similar spatial organization has been observed in self-propelled rod simulations using either rigid~\cite{r32} or elastic~\cite{starruss} elongated particles. 
We found that at high densities the collective dynamics changes and cells organize into vortices. 
This finding cannot be account by the proposed clustering theory, but it is reminiscent of what is observed in self-propelled rod simulations at high densities, though
  in experiments vortices seem to be stable structures while in simulations vortices are unstable. 
From the comparison of these two non-reversing strains, we conclude that  unidirectional cell motion induces the formation of large moving clusters at low and intermediate densities,   
while it results into vortex formation at very high densities. 
On the light of the clustering theory and given the remarkable similarity with self-propelled rod simulations, we suggest that the 
spatial self-organization in these two strains occurs in absence of biochemical signal regulations and as result of the the  combined effect of self-propulsion and volume exclusion interactions. 
All these results strongly suggest that the combination of self-propulsion and steric interaction is a valid pattern formation mechanism which could be also at play in 
recent experiments with {\it Escherichia coli}~\cite{r600} and driven actin filaments~ \cite{r400},  which makes us wonder about the connection between this mechanism and   
 the large body of work on  simple models of self-propelled particles where spontaneous segregation  and long-range orientational has been reported~\cite{r212,r213,r213b,r214,r215,r220,r225,r235,r240}.  

The study of  wild-type cells has revealed that cell reversal affects dramatically the collective dynamics. 
We found that wild-type cells exhibit  cluster size distributions exponentially distributed at low and intermediate densities. 
On the other hand, we measured an the scaling of the cluster perimeter with cluster size characterized by a large exponent $\omega \sim 0.8$ which indicates that clusters are strongly elongated 
with comparison to those found in experiments with the two non-reversing mutants.  
Finally, we observed that at high densities cells self-organize into a mesh-like structure.  
A qualitative understanding of this macroscopic behavior is still missing. 
The comparison of the two non-reversing strains and wild-type cells suggest suggests
 that by only switching on and off  the reversal,  cells can modify dramatically their collective behavior, with the suppression of reversal leading 
to collective motion in the form of moving clusters and vortex formation at high densities. 
We note that this observation is consistent with the observed decrease in reversal frequency in the wild-type upon nutrient depletion, which is followed by the formation of large moving clusters and aggregation of cells.

At a more speculative level, our results suggest that the cell density and the rod  shape of the cells may play an essential role  for bacteria to achieve collective motion~\cite{r180, r190}. 
According to self-propelled rod simulations,  an elongated cell shape strongly facilitates collective motion by promoting the formation of larger clusters. 
Another hint that the rod-shape of the moving bacteria is important for collective motion is provided by  the empirical observation that
 many bacteria undergo a dramatic elongation of their cell shape before assembling into larger groups, e.g.
 in {\it Vibrio parahaemolyticus}~\cite{r200}  or {\it B. subtilis}~\cite{r210}.
Finally, the reported results increase the plausibility of earlier biological hypotheses ~\cite{r180}, that multicellular organization may 
be achieved by  regulating the cell density   via proliferation and cell length  
by direct developmental control. 

We acknowledge support by the German Ministry for Education and Research (BMBF) through grants no. 0315259 and no. 0315734, by the Human Frontier Science Program (HFSP) through grant no. RGP0016-2010, by DFG through grant DE842/2 and the Max Planck Society. Partial DFG support by SFB 555 and GRK 1558 is also acknowledged. AD is a member of the DFG Research Center for Regenerative Therapies Dresden -- Cluster of Excellence -- and gratefully acknowledges support by the Center. FP acknowledges support by PEPS PTI ``Anomalous fluctuations in the collective motion of self-propelled particles".

\section{Material and methods}

\subsection{Bacterial strains.} 
The fully motile strain DK1622 (\wildtype) was used as a wild type~\cite{sup_100} and all other strains used are derivatives of DK1622.
The non-reversing strain DK8505~\cite{sup_200} is referred to as \asmotile.
To generate SA2407 cells, here referred to as \amotile, the {\it frz}
loss-of-function allele $frzCD$::$Tn5$ $lac$ $\Omega$536 from DK8505~\cite{sup_200} was introduced into the $\Delta$$pilA$ 
strain DK10410~\cite{sup_300}, which is unable to assemble type IV pili, using standard procedures~\cite{sup_400}. 
To generate SA2082 ($\Delta$$pilA$, $romR$::$nptII$), the non-motile {\it M. xanthus} mutant referred to as \nonmotile, 
the $romR$::$nptII$ loss-of-function allele from SA1128~\cite{sup_500} was introduced into DK10410. 
All strains used had a doubling time of approximately 5 hrs in CTT liquid medium at $32^{\circ}{\rm C}$. 
Notice that the relaxation time of spatial patterns is below 120 min which implies that the doubling time has a weak effect on the spatial patterns.

\subsection{Cluster formation experiments} 
Cultures of {\it M. xanthus} were grown in CTT liquid medium~\cite{sup_600} at $32^{o}C$ with shaking to an estimated density of
$7\times10^9$ cells/ml. 
Subsequently, cells were diluted to densities of $0.5\times10^8$/ml, $1.0\times10^8$/ml, $1.5\times10^8$/ml, $1.75\times10^8$/ml 
and $2\times10^8$/ml, respectively. 
Cell densities were confirmed by colony counts on CTT agar plates manually and by counting the number of cells 
using a counting chamber manually. 
$30\mu l$ aliquots of cells were transferred to a microscope slide covered with a $1.0$\% agar pad in $0.5$\% 
CTT medium. 
The time point at which the cell drop was completely absorbed in the agar was set as $t = 0$.
For each cell density, 16 slides were prepared and every $30$min (starting at $30$min) up to $480$min, 
a sample was analysed by microscopy using a Leica DM6000B microscope with a Leica $20\times$ phase-contrast 
objective and imaged with a Leica DFC 350FX camera. 
20 phase-contrast images were taken at $20\times$ magnification across a spot. 
After 480min a short time-lapse movie was taken to verify that cells and clusters were migrating. 
%
%
%
%
\begin{figure*}
\centering\resizebox{12cm}{!}{\rotatebox{0}{\includegraphics{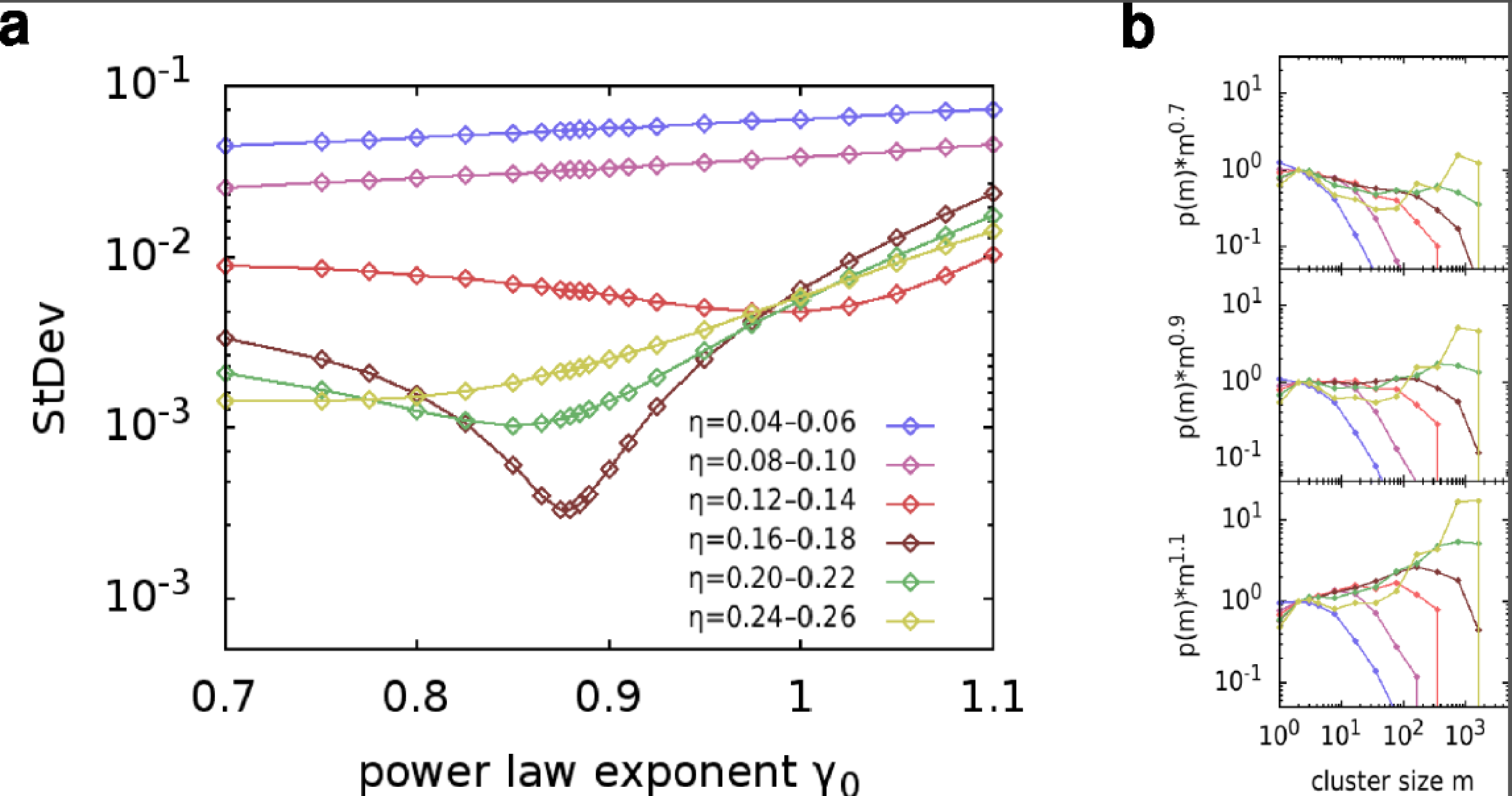}}}
\caption{
Procedure to estimate the critical exponent $\gamma_0$ and critical packing fraction $\eta_c$. a) The standard deviation of the transformation given by Eq.~(\ref{eq:trans}) with respect to its mean value $W$, see Eq.~(\ref{eq:sigma}). The shown data corresponds to \amotile cells. The minimum exhibited by $\eta=0.16-0.18$ indicates that the transformation given by Eq.~(\ref{eq:trans}) leads to a horizontal line, indirectly showing that the values of $\eta$ can be well fitted by 
Eq. (\ref{eq:scaling}). b) shows the sensitivity of this procedure. If the critical exponent is either overestimated or underestimated, there is no transformed CSD, for any value of $\eta$, yielding a horizontal line. Only very close to the actual critical exponent $\gamma_0$, the transformation can be approximated by a constant $W$.}\label{fig:CalcGamma}
\end{figure*}

\subsection{Image analysis} 
Clustering images were taken at $20\times$ magnification. 
Images contain cell clusters as dark regions, often surrounded by a light halo.
%
Cluster boundaries were detected in a multi-step processing queue. 
After initial image normalization, edge detection via the Canny-Deriche algorithm was applied for two 
different levels of spatial detail. 
Both edge images were superimposed subsequently. 
Next, edges were filtered out that surround halos and other non-cluster objects. 
Finally, all incomplete detections were revised/corrected manually in a post-processing step. 
%
%
The areas of the clusters in pixels were extracted using an implementation of the processing queue in the image 
processing tool ImageJ (http://rsbweb.nih.gov/ij/). 
The number of cells inside a cluster, i.e., the cluster size, was estimated as the area of a cluster divided 
by the mean area covered by a single cell, which was found to be 150 pixels at 20-fold magnification. 
According to this definition, a cluster is a connected group of cells, regardless of their orientation.
Packing fraction estimates per image were obtained as the ratio of area covered by cells 
 and the whole area of the image ($1392\times1040$ pixels 
corresponding to  $699 \mu m \times 522 \mu m$).

\subsection{Statistical analysis} 
After applying the image analysis procedure described above to a given image $I$, 
corresponding to a given packing fraction, a large array of various cluster sizes 
is obtained, and $n_I (m,t)$ can be computed. 
We represent by $n_I(m,t)$ the number of clusters of size $m$ in the image I. 
To build the CSD we make use of all the available images corresponding to the given packing fraction $\eta$. 
Let the auxiliary function $g_I(m,t)$ be $g_I(m,t)=m\,n_I(m,t)$. The average of this function reads:
\begin{equation} \label{eq:gm}
g(m,t) = (1/M) \sum_{I} g_I(m,t) \,,
\end{equation}
where $M$ is the number of available images. 
To cope with the sparseness of the data for large cluster 
sizes we implemented several binning procedures, in particular, linear and exponential binning. 
In the following we explain the exponential binning procedure. 
The cluster size space is divided into bins, the first bin contains all clusters of size $s$, $0<s \leq 1$, 
the second bin all clusters of size $1<s \leq 2$, the third bin, $2<s \leq 4$,... the $n$-th bin 
contains cluster sizes $2^{n - 1} < s  \leq 2^n$. 
It is useful to define the function:
\begin{equation} 
g_{bin}(n,t) = \sum_{e(n)}^{e(n+1)} \sum_{I} g_I(m,t) \,
\end{equation}
where $e(n)= 2^n$. 
The binned CSD is defined as: 
\begin{equation} \label{eq:pbin}
p_{bin}(e(n),t) = \frac{g_{bin}(n,t)}{C  (e(n-1)-e(n))} \,.
\end{equation}
Thus, $\sum_m p_{bin}(m=e(n),t) (e(n-1)-e(n)) =1$. It is worth noticing that if the underlying 
CSD $p(m)$ is a power-law characterized by an exponent $\gamma$, i.e., $p(m) \sim m^{-\gamma}$, the exponential 
binning procedure given by Eq. (\ref{eq:pbin}) results in $p_{bin}(m) \sim m^{-\gamma}$. 
On the other hand, if the underlying CSD $p(m)$ is an exponential, i.e., $p(m) \sim \exp(m/m_0)$, the exponential binning 
leads to $p_{bin}(m) \sim m^{-1} \exp(m/m_0)$. 
In the text, for simplicity we referred to $p_{bin}(m,t)$ just as $p(m,t)$.

In what follows, we explain how the critical exponent has been measured. At the critical packing fraction $\eta_c$ the CSD is a power-law (with an exponential cut-off due to the finite number of cells). The problem consists in identifying the critical packing fraction $\eta_c$ and the critical exponent $\gamma_0$. Assuming that we know $\gamma_0$ at $\eta_c$ the following transformation yields to a constant:
\begin{equation}\label{eq:trans}
y(m)=p(m;\eta_c)m^{\gamma}= W
\end{equation}
where $W$ is a constant and the equality holds true for $1<x<x_{cut-off}$, where $x_{cut-off}$ denotes the beginning of the cut-off. The value of W is the average value of $y(m)$ in the interval $1<x<x_{cut-off}$. This means that if we plot $y(m)$  vs. $m$, we observe a horizontal line at the critical packing fraction $\eta_c$. We can measure how close we are to the horizontal line by computing: 
\begin{equation}\label{eq:sigma}
\sigma^2(\eta, \gamma) = \sum (y(m)-W)^2 
\end{equation}
By minimizing Eq.~(\ref{eq:sigma}) with respect to $\eta$ and $\gamma$, the critical packing fraction and critical exponent can be obtained. Figure \ref{fig:CalcGamma} illustrates the procedure. In the figure the cut-off was taken to $x_{cut-off}=220$ (various other values were also studied). We found that the critical packing fraction lays between $0.16$ and $0.18$ for either \amotile or \asmotile cells and the critical exponent is $\gamma_0=0.88 \pm 0.07$ and $\gamma_0=0.85 \pm 0.07$ for \amotile or \asmotile cells, respectively.

\end{document}